%% file: iclr2024_conference.tex
\documentclass{article} 
\usepackage{iclr2024_conference,times}

\input{math_commands.tex}

\usepackage[utf8]{inputenc}
\usepackage{hyperref}
\usepackage{url}
\usepackage{graphicx} 
\usepackage{caption}
\usepackage{subcaption}
\usepackage{multirow}
\usepackage{float}
\iclrfinalcopy
\title{Audio Processing using Pattern Recognition for
Music Genre Classification}

\author{Sivangi Chatterjee, Srishti Ganguly, Avik Bose, Hrithik Raj Prasad, Arijit Ghosal  \\
Department of Information Technology\\
St. Thomas' College of Engineering \& Technology, MAKAUT\\
\texttt{Email: \{sivangichatterjee999, arijitghosal.stcet\}@gmail.com}
}

\begin{document}

\maketitle

\begin{abstract}


This project explores the application of machine learning techniques for music genre classification using the GTZAN dataset, which contains 100 audio files per genre. Motivated by the growing demand for personalized music recommendations, we focused on classifying five genres—Blues, Classical, Jazz, Hip Hop, and Country—using a variety of algorithms including Logistic Regression, K-Nearest Neighbors (KNN), Random Forest, and Artificial Neural Networks (ANN) implemented via Keras. The ANN model demonstrated the best performance, achieving a validation accuracy of 92.44\%. We also analyzed key audio features such as spectral roll-off, spectral centroid, and MFCCs, which helped enhance the model’s accuracy. Future work will expand the model to cover all ten genres, investigate advanced methods like Long Short-Term Memory (LSTM) networks and ensemble approaches, and develop a web application for real-time genre classification and playlist generation. This research aims to contribute to improving music recommendation systems and content curation.
\end{abstract}

\section{Introduction}
With the onset of the 21st century, the music landscape has expanded dramatically, leading to a surge in genre diversity. Nowadays, people often explore music by genres rather than focusing solely on individual artists or bands. The digital age has opened up a vast range of musical styles—from classical to modern innovations—making genre classification increasingly challenging. With music catalogs expanding
rapidly, manually categorizing songs into their respective genres is no longer feasible. This is where technology plays a crucial role. Platforms such as Spotify and SoundCloud leverage music classification technologies to provide precise recommendations to users, while apps like Shazam focus on music recognition and classification, reflecting the growing dependence on auto-
mated systems in the music industry. \citep{Content_JHao} \citep{towardsdatascienceMusicGenre}

Machine learning and classification algorithms have become indispensable in this context. These technologies analyze audio data to detect patterns and trends, enabling accurate genre classification. By assessing features like rhythm, tempo, instrumentation, and vocal style, machine learning models can reliably identify a song’s genre. This enhances user recommendations and drives innovation in music discovery and content curation. \citep{1021072}

This project aims to develop an advanced machine learning model to classify music genres by extracting and analyzing relevant audio features. A general machine learning block diagram is illustrated in Figure 1. It starts with data acquisition from various sources, followed by data preprocessing, where raw data is filtered to improve its quality. The preprocessed data is then used to extract relevant features, and the dataset is divided into training and evaluation subsets. After selecting a suitable machine learning model based on the characteristics of the data, the model is trained using the training subset. Subsequently, the model's performance is tested with the evaluation subset to measure accuracy and precision. Based on these results, the model is adjusted and fine-tuned until it meets the desired standards for deployment.

This initiative seeks to advance music classification technology, for navigating the extensive and varied realm of contemporary music, significantly enhancing the ability to categorize and recommend music effectively.

\begin{figure}[htbp]
\centering\includegraphics[width=\textwidth]{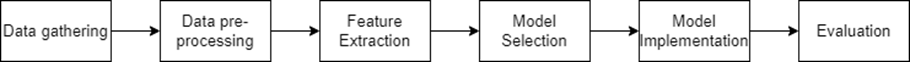}
    \caption{Machine Learning Block Diagram}
    \label{fig:evaluation_schema}
\end{figure}

\section{Literature Survey}
Music genre classification has been a prominent topic in audio signal processing, with several studies contributing unique methodologies. Sanghamitra Das, Suchibrota Dutta, and Dr. Arijit Ghosal (2020) focus on the stratification of Indian dance forms through audio signals, using feature extraction techniques similar to mine. However, their focus on Indian classical music contrasts with my broader exploration of global music genres. Tzanetakis and Cook (2002) established the foundational framework for genre classification using spectral features like MFCCs and ZCR, which I also employ. However, my research advances their work by incorporating deep learning models, specifically convolutional neural networks (CNNs), as demonstrated by Vishnupriya and Meenakshi (2018). While their CNN-based approach aligns with my work, I further refine it by combining multiple machine learning techniques for more accurate and robust genre classification. In conclusion, while these studies laid important groundwork, my research extends the field by integrating traditional and modern methods to improve genre classification accuracy across a broader range of music.\citep{inbook} \citep{1021072} \citep{Vishnupriya2018AutomaticMG}

\section{Design Methodology}
\label{section:design methodology}

\subsection{Dataset} 

The dataset employed in this project is the GTZAN Genre Collection, created by George Tzanetakis. This dataset originates from MARSYAS, an open-source software framework known for its capabilities in audio processing and Music Information Retrieval (MIR). MARSYAS, which stands for Music Analysis, Retrieval, and Synthesis for Audio Signals, is widely used in the field of audio analysis.\citep{Lidy2008}

The GTZAN Genre Collection comprises 1,000 audio tracks, each with a duration of 30 seconds. The dataset comprises 10 distinct music genres, including  Disco,  Pop, Classical, Blues, Country, Jazz, Metal, Hip-Hop, Reggae, and Rock. Each genre is represented by a collection of 100 tracks. All tracks are stored as 22050 Hz Mono 16-bit audio files in the .wav format.

This well-structured dataset provides a balanced representation of various musical genres, making it suitable for training machine learning models for genre classification tasks. Its origins from MARSYAS ensure that it is a valuable resource for advancing research in audio signal processing and music information retrieval.\citep{Elbir,clairvoyantMusicGenre}

\subsection{Feature Extraction}
\label{feature-extraction}
Sound is characterized as an audio signal that can be described by parameters like bandwidth, frequency and volume in decibels. These signals can be modeled as a function of amplitude over time. In this project, sound samples are stored as digital audio files in the .wav format.

The initial phase of audio signal processing involves feature extraction, which is essential for sound analysis. Features such as spectral characteristics, chroma features, and Mel-Frequency Cepstral Coefficients (MFCCs) are commonly extracted, as they capture the essential properties of audio signals relevant to tasks like classification and recognition. Spectral features, for instance, represent the frequency content of the signal, while MFCCs offer a compact representation of the spectral envelope, which is particularly useful in speech and music analysis.

After extracting these features, they serve as the foundation for various decision-making processes, including detection, classification, and knowledge integration. Python libraries like librosa offer powerful tools for extracting these audio features, making complex audio processing tasks more manageable and efficient for machine learning applications. 

\subsection{Feature Selection} 
\label{subsection:feature-selection}
Each audio signal encompasses a wide range of features that can be extracted for analysis. However, to tackle a specific problem effectively, it's crucial to concentrate on the most pertinent features. This targeted approach helps prevent overfitting, enhances model accuracy, and shortens training time by removing unnecessary complexity. 

Feature selection plays a vital role in machine learning by filtering out irrelevant or redundant data that might introduce noise into the model. By prioritizing the most important features, the model is better optimized for both performance and efficiency. The key features selected for this project are described below. 

\subsubsection{Zero Crossing Rate(ZCR)}
\label{subsubsection:zcr}



The Zero Crossing Rate (ZCR) measures how frequently an audio signal transitions through the zero amplitude point, indicating changes between positive and negative values. Mathematically, it is expressed as:

\[
ZCR = \frac{1}{T-1} \sum_{t=1}^{T-1} \mathbf{1}(x_t \cdot x_{t+1} < 0)
\]

where \( T \) is the total number of samples, \( x_t \) is the signal value at time \( t \), and \( \mathbb{1} \) is an indicator function that equals 1 when there is a sign change between consecutive samples, indicating a zero crossing, and 0 otherwise. The result is normalized between 0 and 1, with higher ZCR values corresponding to more frequent zero crossings, and thus higher frequency content in the signal. ZCR is instrumental in distinguishing between different audio types. It helps in identifying silence, noise, and structured audio like speech or music. For example, in music genre classification, genres like metal and rock often exhibit higher ZCR values due to their rapid frequency shifts. Therefore, ZCR is a critical feature for distinguishing between audio content and non-audio elements like silence or noise, as well as for differentiating between speech and background sounds.

Given its ability to highlight frequency changes and differentiate audio types, ZCR is an important feature for this project. Its use improves the model's ability to classify and analyze audio signals across various music genres effectively.
\citep{Kedem1986}

\begin{figure}[!ht]
    \centering
    \begin{subfigure}[b]{0.45\textwidth}
        \centering
        \includegraphics[width=\textwidth]{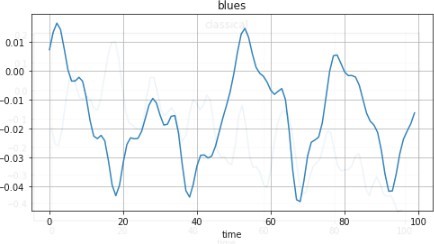}
        \caption{Blues}
    \end{subfigure}
    \hfill
    \begin{subfigure}[b]{0.45\textwidth}
        \centering
        \includegraphics[width=\textwidth]{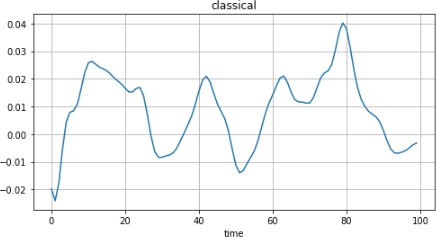}
        \caption{Classical}
    \end{subfigure}

    \vskip\baselineskip

    \begin{subfigure}[b]{0.45\textwidth}
        \centering
        \includegraphics[width=\textwidth]{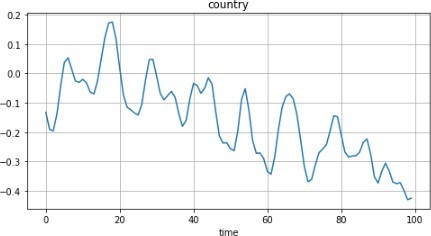}
        \caption{Country}
    \end{subfigure}
    \hfill
    \begin{subfigure}[b]{0.45\textwidth}
        \centering
        \includegraphics[width=\textwidth]{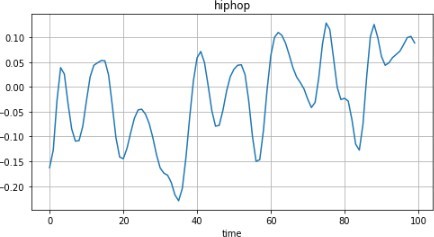}
        \caption{Hip-Hop}
    \end{subfigure}
    
    \vskip\baselineskip

    \begin{subfigure}[b]{0.45\textwidth}
        \centering
        \includegraphics[width=\textwidth]{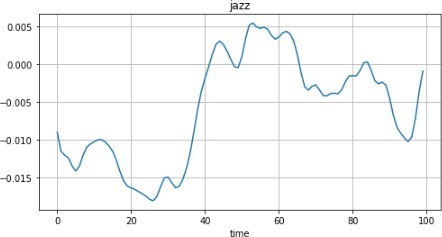}
        \caption{Jazz}
    \end{subfigure}

    \caption{ZCR variation across five genres}
    \label{fig:grid_example}
\end{figure}
\newpage
\subsubsection{Spectral Centroid}
\label{subsubsection:spectral centroid}
The spectral centroid measures where the center of gravity lies in the frequency distribution of a sound. It reflects the average frequency of the audio signal, weighted by the amplitude of each frequency component.  It is mathematically defined as the sum of the product of each frequency ($f_k$) and its corresponding amplitude ($A_k$), divided by the sum of all amplitudes. This can be expressed with the formula:

\[
\text{Spectral Centroid} = \frac{\sum_{k=1}^{N} f_k \cdot A_k}{\sum_{k=1}^{N} A_k}
\]

In this formula, $f_k$ represents the frequency of the $k$-th component, $A_k$ is the amplitude of that component, and $N$ is the total number of frequency components in the signal. 

When analyzing different music genres, such as blues versus metal, the spectral centroid can highlight distinctive features. Blues music, with its stable and consistent frequency content, often has a spectral centroid that is centrally located within the frequency range. On the other hand, metal music, which is characterized by a broader and more varied frequency range, tends to have its spectral centroid positioned towards the higher or lower extremes. This distinction in the spectral centroid helps in differentiating music genres based on their frequency characteristics. \citep{LiTao}
\begin{figure}[!ht]
    \centering
    \begin{subfigure}[b]{0.45\textwidth}
        \centering
        \includegraphics[width=\textwidth]{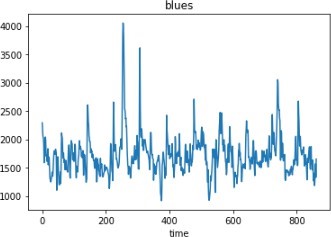}
        \caption{Blues}
    \end{subfigure}
    \hfill
    \begin{subfigure}[b]{0.45\textwidth}
        \centering
        \includegraphics[width=\textwidth]{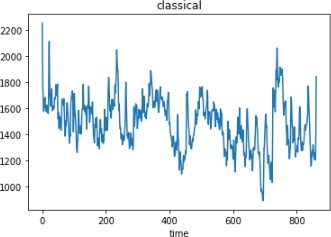}
        \caption{Classical}
    \end{subfigure}

    \vskip\baselineskip

    \begin{subfigure}[b]{0.45\textwidth}
        \centering
        \includegraphics[width=\textwidth]{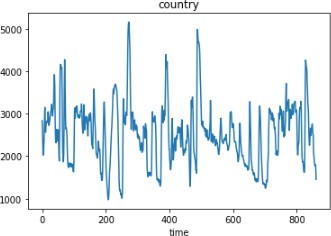}
        \caption{Country}
    \end{subfigure}
    \hfill
    \begin{subfigure}[b]{0.45\textwidth}
        \centering
        \includegraphics[width=\textwidth]{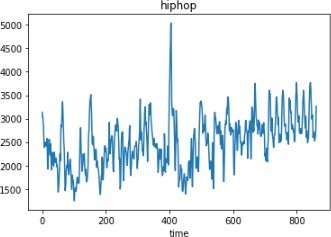}
        \caption{Hip-Hop}
    \end{subfigure}
    
    \vskip\baselineskip

    \begin{subfigure}[b]{0.45\textwidth}
        \centering
        \includegraphics[width=\textwidth]{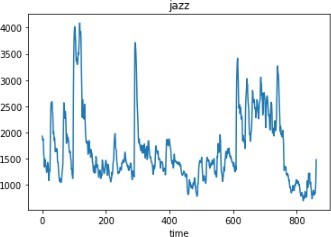}
        \caption{Jazz}
    \end{subfigure}

    \caption{Spectral Centroid variation across five genres}
    \label{fig:grid_example}
\end{figure}

\subsubsection{Spectral Roll-Off}
\label{subsubsection:spectral rolloff}
Spectral roll-off defines the frequency threshold below which a certain percentage of the total energy in an audio signal is concentrated. For example, it might indicate the frequency below which 85\% of the energy of the signal is concentrated. This feature helps us understand how frequencies are distributed within the signal.
This can be calculated using the formula:

\[
\sum_{k=1}^{k_\text{rolloff}} A_k = p \cdot \sum_{k=1}^{N} A_k
\]

where $A_k$ represents the amplitude of the $k$-th frequency bin, $N$ is the total number of frequency bins, $p$ is the percentage (such as 0.85 for 85\%), and $k_\text{rolloff}$ is the frequency bin index at which the cumulative energy reaches the desired percentage.

Different music genres exhibit distinctive spectral roll-off characteristics. In genres like rock or metal, which have a wide frequency range and substantial high-frequency content, the spectral roll-off point is usually higher. This is because a larger portion of the energy is present in the higher frequencies. Conversely, genres such as blues or classical music, which focus on lower frequencies, tend to have a lower spectral roll-off. This characteristic is useful for differentiating and categorizing various types of music based on their frequency profiles.\citep{LiTao}

\begin{figure}[!ht]
    \centering
    \begin{subfigure}[b]{0.45\textwidth}
        \centering
        \includegraphics[width=\textwidth]{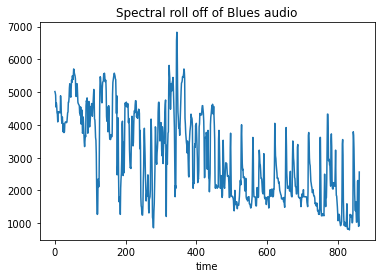}
        \caption{Blues}
    \end{subfigure}
    \hfill
    \begin{subfigure}[b]{0.45\textwidth}
        \centering
        \includegraphics[width=\textwidth]{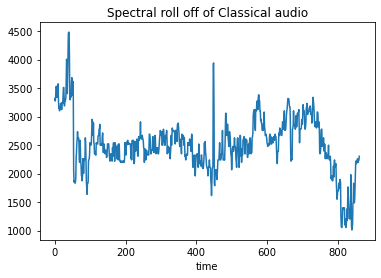}
        \caption{Classical}
    \end{subfigure}

    \vskip\baselineskip

    \begin{subfigure}[b]{0.45\textwidth}
        \centering
        \includegraphics[width=\textwidth]{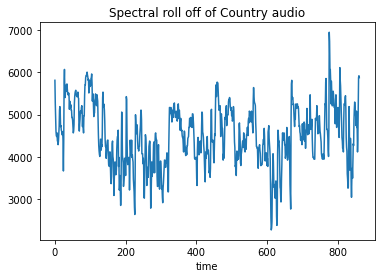}
        \caption{Country}
    \end{subfigure}
    \hfill
    \begin{subfigure}[b]{0.45\textwidth}
        \centering
        \includegraphics[width=\textwidth]{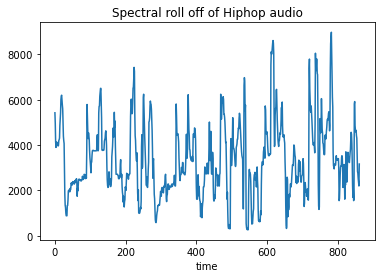}
        \caption{Hip-Hop}
    \end{subfigure}
    
    \vskip\baselineskip

    \begin{subfigure}[b]{0.45\textwidth}
        \centering
        \includegraphics[width=\textwidth]{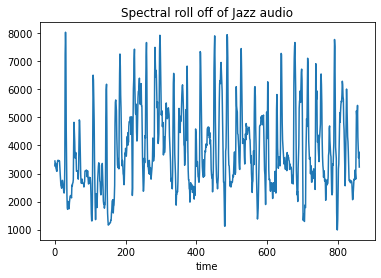}
        \caption{Jazz}
    \end{subfigure}

    \caption{Spectral Roll-Off variation across five genres}
    \label{fig:grid_example}
\end{figure}

\subsubsection{Mel Frequency Cepstral Coefficients(MFCC)}
\label{subsubsection:mfcc}
Mel-Frequency Cepstral Coefficients (MFCCs) are a group of features, typically ranging from 10 to 20, that characterize the general contour of an audio signal's spectral envelope. 
These coefficients are designed to reflect how the human ear perceives sound, making them especially effective for analyzing the acoustic properties of audio signals. The process of computing MFCCs involves several steps. First, the Fourier transform of the signal is taken to obtain its power spectrum. Then, a Mel filter bank is applied to emphasize frequencies that are more perceptually important to human hearing, with the Mel scale defined as:

\[
\text{Mel}(f) = 2595 \cdot \log_{10}\left(1 + \frac{f}{700}\right)
\]

After applying the filter bank, the logarithm of the Mel-filtered signal is taken to model human loudness perception, followed by a discrete cosine transform (DCT) to generate the MFCCs. This process provides a detailed representation of the spectral characteristics of the signal, making MFCCs particularly useful in tasks like music analysis and speech recognition, where understanding how humans perceive sound is crucial.
\citep{West2004} \citep{inbook}

\begin{figure}[!ht]
    \centering
    \begin{subfigure}[b]{0.45\textwidth}
        \centering
        \includegraphics[width=\textwidth]{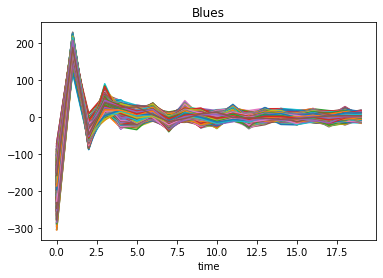}
        \caption{Blues}
    \end{subfigure}
    \hfill
    \begin{subfigure}[b]{0.45\textwidth}
        \centering
        \includegraphics[width=\textwidth]{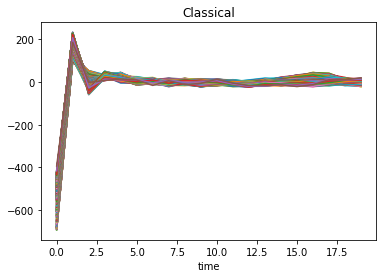}
        \caption{Classical}
    \end{subfigure}

    \vskip\baselineskip

    \begin{subfigure}[b]{0.45\textwidth}
        \centering
        \includegraphics[width=\textwidth]{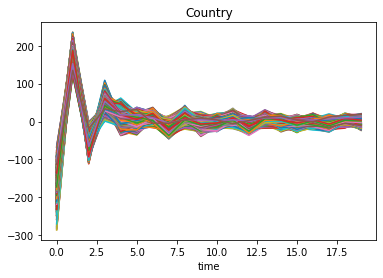}
        \caption{Country}
    \end{subfigure}
    \hfill
    \begin{subfigure}[b]{0.45\textwidth}
        \centering
        \includegraphics[width=\textwidth]{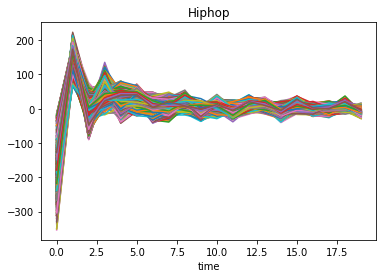}
        \caption{Hip-Hop}
    \end{subfigure}
    
    \vskip\baselineskip

    \begin{subfigure}[b]{0.45\textwidth}
        \centering
        \includegraphics[width=\textwidth]{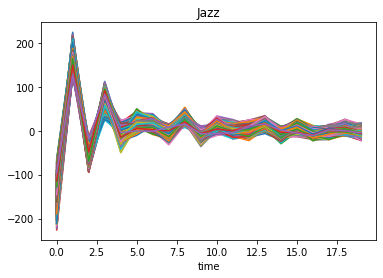}
        \caption{Jazz}
    \end{subfigure}

    \caption{MFCC variation across five genres}
    \label{fig:grid_example}
\end{figure}

\subsubsection{Chroma Shift}
\label{subsubsection:chroma shift}
Chroma features provide a powerful method for representing musical audio by converting the entire frequency spectrum into 12 bins, each corresponding to one of the 12 semitones in an octave. These features capture the harmonic and melodic aspects of music by focusing on pitch, while being robust to variations in timbre and instrumentation. The chroma feature extraction process involves computing the short-time Fourier transform (STFT) of the signal, followed by grouping the resulting frequencies into 12 chroma bins using a mapping function such as:

\[
C(f) = \text{mod}\left(12 \cdot \log_2\left(\frac{f}{f_\text{ref}}\right), 12\right)
\]

where $f$ is the frequency and $f_\text{ref}$ is a reference frequency. This transformation emphasizes musical pitch while reducing sensitivity to the specific instruments or timbres involved, making chroma features particularly valuable for tasks like music analysis and genre classification, as they capture the fundamental musical elements of a piece. \citep{clairvoyantMusicGenre}


\begin{figure}[!ht]
    \centering
    \begin{subfigure}[b]{0.40\textwidth}
        \centering
        \includegraphics[width=\textwidth]{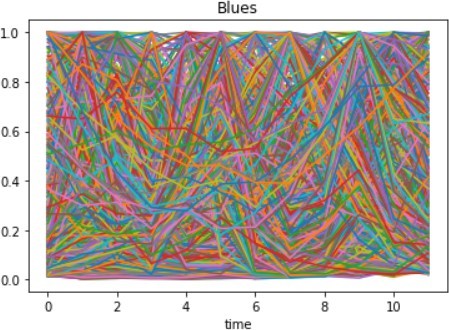}
        \caption{Blues}
    \end{subfigure}
    \hfill
    \begin{subfigure}[b]{0.40\textwidth}
        \centering
        \includegraphics[width=\textwidth]{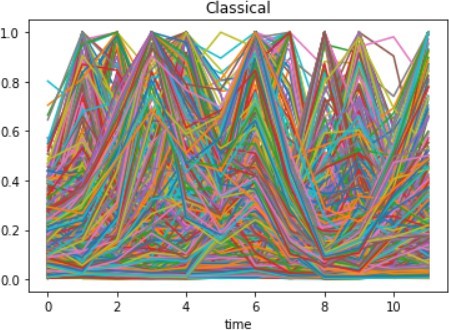}
        \caption{Classical}
    \end{subfigure}

    \vskip\baselineskip

    \begin{subfigure}[b]{0.40\textwidth}
        \centering
        \includegraphics[width=\textwidth]{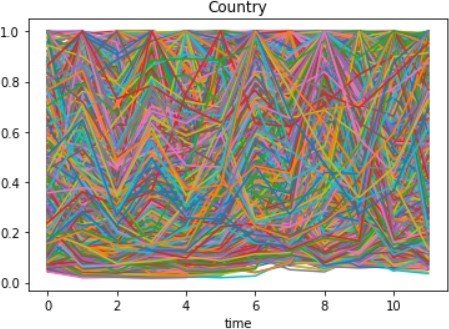}
        \caption{Country}
    \end{subfigure}
    \hfill
    \begin{subfigure}[b]{0.40\textwidth}
        \centering
        \includegraphics[width=\textwidth]{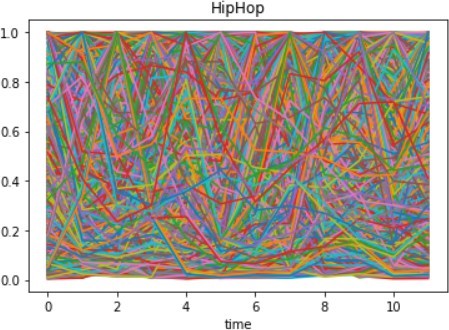}
        \caption{Hip-Hop}
    \end{subfigure}
    
    \vskip\baselineskip

    \begin{subfigure}[b]{0.40\textwidth}
        \centering
        \includegraphics[width=\textwidth]{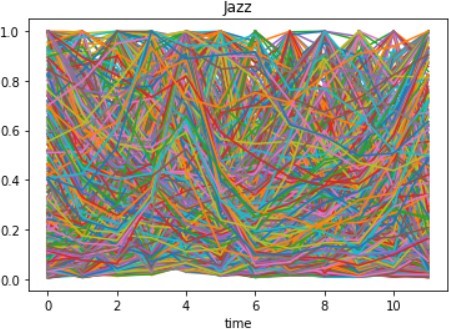}
        \caption{Jazz}
    \end{subfigure}

    \caption{Chroma Shift variation across five genres}
    \label{fig:grid_example}
\end{figure}
\newpage
\subsubsection{Data Processing}
\label{subsubsection:data pre-processing}
Before the raw data could be used for training, it underwent extensive preprocessing to correct distortions and address various errors, including radiometric inaccuracies. A critical step in this process involved scaling the features to ensure consistency across the dataset, bringing all features to a uniform scale. This was achieved using techniques such as the Standard Scaler and Min-Max Scaler from the Scikit-learn library. Additionally, to maintain consistency in audio data, all audio files were converted to a 16-bit mono format using Audacity, ensuring compatibility and uniformity across the dataset for further analysis and model training.
\section{Classification Models}

In machine learning, classification focuses on predicting a class label for a given input. This process involves training a model with data to recognize patterns and relationships that link input features to class labels. To achieve good performance, the training data should be diverse and accurately reflect the problem at hand, with sufficient examples for each class to minimize bias and enhance generalization. The effectiveness of a classification model also depends on the selection of features and preprocessing methods. The following sections will explore the classification models used in this project, detailing their strengths and specific use cases. \citep{PRML,PC}

\label{section:classification}
\subsection{K-nearest Neighbours}
The k-Nearest Neighbors (KNN) algorithm is a straightforward and powerful technique employed in supervised machine learning for classification tasks. It classifies a data point based on the most common class among its k nearest neighbors in the training dataset.  For instance, when k  is set to 1, the data point is classified according to the class of its closest neighbor.

KNN is non-parametric, meaning it doesn't assume any specific data distribution, which allows it to be versatile across different scenarios. However, the choice of k and the distance metric (like Euclidean or Manhattan distance) can impact its performance. Additionally, KNN can be computationally expensive, particularly with large datasets, as it requires distance calculations for all training points for each prediction.

Despite these challenges, KNN is often used as a baseline in classification tasks due to its simplicity and interpretability. Proper feature scaling, such as using Standard Scaler or Min-Max Scaler, can enhance its effectiveness by ensuring balanced distance calculations. \citep{Bhat}

\begin{figure}[!ht]
    \centering
    \begin{subfigure}[b]{0.30\textwidth}
        \centering
        \includegraphics[width=\textwidth]{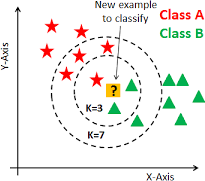}
        \caption{Diagram of KNN classification \citep{mediumKnearestNeighbor}}
    \end{subfigure}
    \hfill
    \begin{subfigure}[b]{0.60\textwidth}
        \centering
        \includegraphics[width=\textwidth]{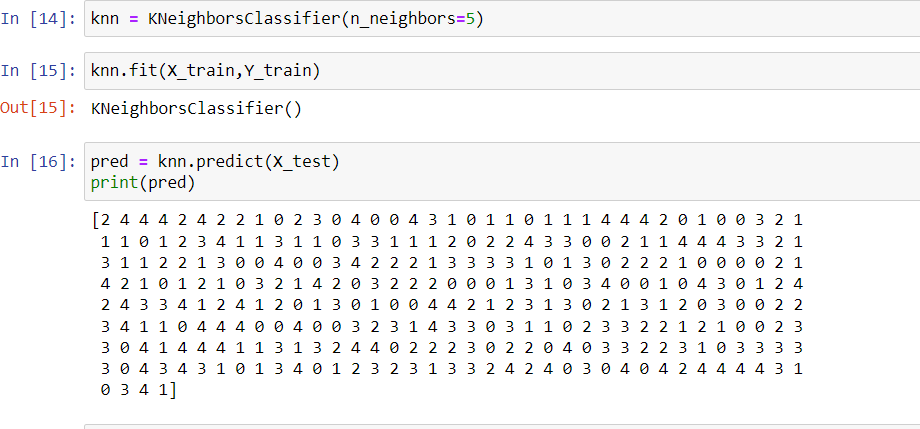}
        \caption{KNN model using scikit-learn}
    \end{subfigure}
\end{figure}

\subsection{Logistic Regression}
Logistic regression is named after the logistic function, which is crucial to the method. The sigmoid function, a key component, converts predicted values into probabilities, mapping any real number to a range between 0 and 1. This results in an S-shaped curve where values on the y-axis are between 0 and 1, with 0.5 as the cutoff point for binary classification.

Once probabilities are predicted, the model's parameters are evaluated by computing the likelihood, which measures how well the observed data aligns with the model's predictions. This involves determining how different parameter values support the observed outcomes. A cost function is then used to quantify the error between the predicted probabilities and the actual outcomes. To minimize this error and refine the model, gradient descent is employed to iteratively update the parameters, striving for the best fit.

\begin{figure}[!h]
 \centering
    \begin{subfigure}[b]{0.40\textwidth}
        \centering
        \includegraphics[width=\textwidth]{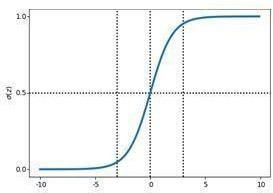}
        \caption{Sigmoid Function \citep{mediumImplementationGradient}}
    \end{subfigure}
    \hfill
    \begin{subfigure}[b]{0.50\textwidth}
        \centering
        \includegraphics[width=\textwidth]{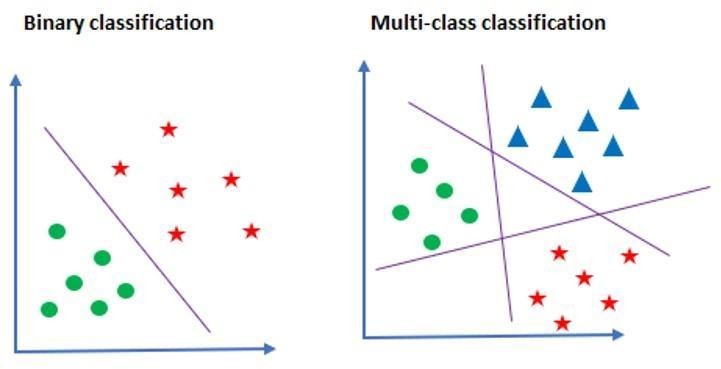}
        \caption{Binary vs MultiClass Classification \citep{mediumMulticlassClassificationOnevsAll}}
    \end{subfigure}

 

\end{figure}

\subsection{Random Forest}
The fundamental idea behind a random forest is to leverage a collection of decision trees operating together to generate predictions. Each decision tree makes an independent prediction, and the final class is determined by selecting the most frequent class among all the trees' predictions.. In our implementation using scikit-learn, the total number of trees in the forest
i.e. estimators = 1000, the splitting evaluation method i.e. criterion is chosen as ‘gini’ and maxdepth is set to 10. \citep{bahuleyan2018musicgenreclassificationusing}

\begin{figure}[!ht]
 \centering
    \begin{subfigure}[b]{0.55\textwidth}
        \centering
        \includegraphics[width=\textwidth]{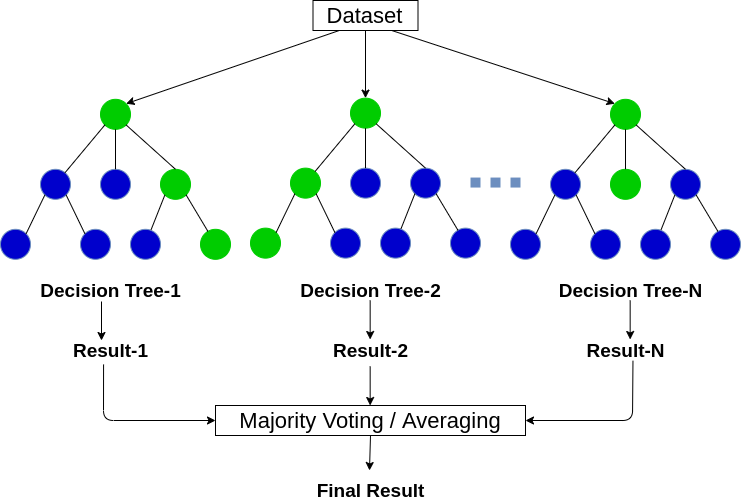}
        \caption{Random Forest \citep{CompleteGuide}}
    \end{subfigure}
    \vskip\baselineskip
    \begin{subfigure}[b]{0.70\textwidth}
        \centering
        \includegraphics[width=\textwidth]{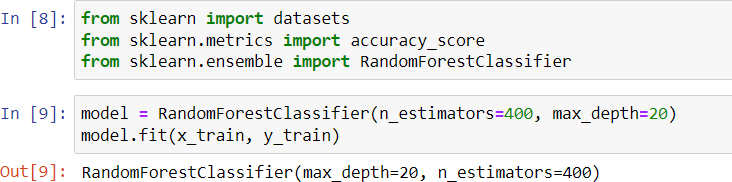}
        \caption{Random Forest Implementation using Scikit}
    \end{subfigure}
\end{figure}

     %
 

\subsection{Artificial Neural Network}
An artificial neural network is like a human brain which consists of several nodes or neurons interconnected like a web. It is composed of an input layer, one or more hidden layers, and an output layer. The input layer receives data in various formats, the hidden layers process and interpret this data, and the output layer generates the final prediction based on the learned information. During the supervised phase the Ann model computes the error of the predicted values from the actual values. Then it uses backpropagation to update the weights and biases of the connections between the processing units until the error value is minimum. Each neuron or processing unit takes an input value, then processes the data using weights and biases and finally passes it through an activation function and converts the input into a more useful output.
The following image shows how a neuron or processing unit works. \citep{Vishnupriya2018AutomaticMG}



\begin{figure}[!ht]
 \centering
    \begin{subfigure}[b]{0.55\textwidth}
        \centering
        \includegraphics[width=\textwidth]{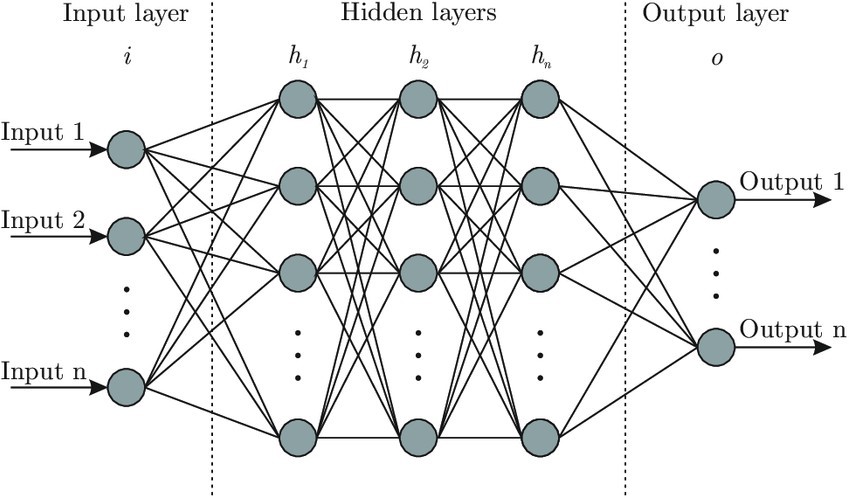}
        \caption{Example of ANN Model \citep{ArtificialNeural}}
    \end{subfigure}
    \vskip\baselineskip
    \begin{subfigure}[b]{0.60\textwidth}
        \centering
        \includegraphics[width=\textwidth]{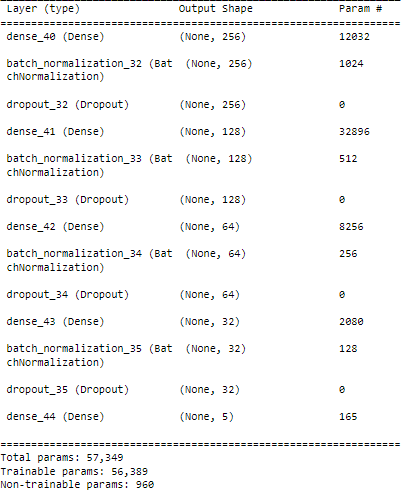}
        \caption{ANN using Keras}
    \end{subfigure}
\end{figure}

\section{Testing}
The GTZAN dataset contains 100 audio files for each genre. In this study, we focused on five genres: Blues, Classical, Jazz, Hip Hop, and Country. For each model, we calculated both the training and validation accuracies to assess their performance. The model with the highest validation accuracy was selected for further testing. This selected model was then evaluated using external, previously unseen audio files, which were not part of the GTZAN dataset, to measure its generalization capability. The following table presents a comparative analysis of the accuracy achieved by various classifiers.

\begin{table}[ht]
    \centering
    \begin{tabular}{|p{4cm}|p{4cm}|p{4cm}|} 
        \hline
        \textbf{Classifier} & \textbf{Segment} & \textbf{Validation Accuracy} \\
        \hline
        KNN & 30s & 87\% \\
        Logistic Regression & 30s & 86\% \\
        Random Forest & 30s & 89\% \\
        Artificial Neural Network & 30s & 92.44\% \\
        \hline
    \end{tabular}
    \caption{Comparative analysis of accuracy achieved by different classifiers.}
    \label{tab:classifier_accuracy}
\end{table}

Based on the results presented in the table, it is clear that the ANN achieved the highest accuracy among the classifiers. The following image illustrate the confusion matrix, which compares the actual music genre classes against the predicted classes for the Keras ANN model. This visualization provides insight into the model's performance across different genres.
\begin{figure}[!ht]
 \centering
 \includegraphics[width=0.75\textwidth]{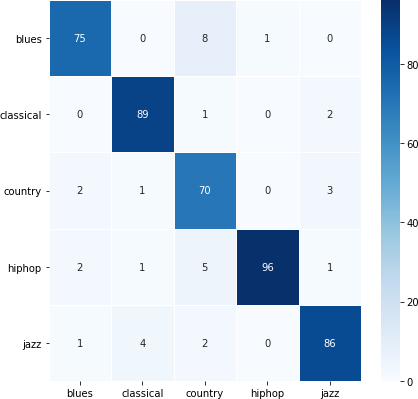}
 \caption{Confusion Matrix fot the ANN Model}
 \label{fig:confusion_matrix}
\end{figure}
\section{Future Planning}
\subsection{Scope for Future Improvement}
This project successfully classified music into five genres, but there is potential for further enhancement by extending the model to handle all ten genres present in the dataset. Future work should focus on developing and optimizing models to improve accuracy across a broader range of genres.

Additionally, exploring advanced techniques such as Long Short-Term Memory (LSTM) networks could substantially boost classification accuracy. LSTM networks excel at handling sequential data, which might improve the model's capacity to identify temporal patterns in audio signals, potentially resulting in more precise genre classification.

Another promising direction is the development of an ensemble model that integrates multiple classification algorithms. By aggregating predictions from various models, such an ensemble approach could leverage the strengths of each individual classifier, leading to more robust and accurate results.

Furthermore, the creation of a web application to streamline the genre classification process is an exciting opportunity. This application could allow users to upload audio files and receive genre predictions in real-time. An additional feature could include generating personalized playlists based on the predicted genre, providing a practical and user-friendly application of the classification model.

These advancements will not only improve classification accuracy but also enhance the overall user experience, making the system more versatile and accessible.

\subsection{Conclusion}

This project illustrates how machine learning can be utilized for music genre classification, an area of significant importance in audio processing and music information retrieval. By focusing on five distinct genres, we employed a range of classification techniques, including K-Nearest Neighbors (KNN), Logistic Regression, Random Forest, and Artificial Neural Networks using the Keras framework. The results show varying degrees of success across these models, with notable performance in accuracy. The insights gained from this project highlight the effectiveness of machine learning in genre classification and provide a foundation for future work that could expand to additional genres or incorporate more complex features. Overall, the project achieves its goals of genre categorization and contributes to the growing field of automated music classification.



\bibliography{iclr2024_conference}
\bibliographystyle{iclr2024_conference}

\end{document}

%% file: math_commands.tex
\usepackage{amsmath,amsfonts,bm}

\def\eqref#1{equation~\ref{#1}}

\def\1{\bm{1}}

\DeclareMathAlphabet{\mathsfit}{\encodingdefault}{\sfdefault}{m}{sl}
\SetMathAlphabet{\mathsfit}{bold}{\encodingdefault}{\sfdefault}{bx}{n}

